\documentclass[amsmath,superscriptaddress,showpacs,prb,twocolumn]{revtex4-1}
\usepackage{bbm}
\usepackage{bm}
\usepackage{enumerate}
\usepackage{graphicx}
\usepackage[dvips]{epsfig}
\usepackage{epsf}
\usepackage[latin1]{inputenc}
\usepackage{amsmath}
\usepackage{amsfonts}
\usepackage{amssymb}
\usepackage{xcolor}
\bibpunct{[}{]}{,}{n}{}{}

\makeatletter
\newcommand{\Rmnum}[1]{\expandafter\@slowromancap\romannumeral #1@}
\makeatletter

\begin{document}
\title{Intrinsic spin Nernst effect of magnons in a noncollinear antiferromagnet}
\author{Bo Li}
\affiliation{Department of Physics and Astronomy and Nebraska Center for Materials and Nanoscience, University of Nebraska, Lincoln, Nebraska 68588, USA}

\author{Shane Sandhoefner}
\affiliation{Department of Physics and Astronomy and Nebraska Center for Materials and Nanoscience, University of Nebraska, Lincoln, Nebraska 68588, USA}

\author{Alexey A. Kovalev}
\affiliation{Department of Physics and Astronomy and Nebraska Center for Materials and Nanoscience, University of Nebraska, Lincoln, Nebraska 68588, USA}
\begin{abstract}
We investigate the intrinsic magnon spin current in a noncollinear antiferromagnetic insulator. We introduce a definition of the magnon spin current in a noncollinear antiferromagnet and find that it is in general non-conserved, but for certain symmetries and spin polarizations the averaged effect of non-conserving terms can vanish. We formulate a general linear response theory for magnons in noncollinear antiferromagnets subject to a temperature gradient and analyze the effect of symmetries on the response tensor. We apply this theory to single-layer potassium iron jarosite KFe$_3$(OH)$_6$(SO$_4$)$_2$ and predict a measurable spin current response. 
\end{abstract}
\maketitle

\noindent

In recent years, advances in research on topological properties of electron systems \cite{RevModPhys.82.3045} have encouraged explorations of manifestations of topology in many other systems, e.g., magnonic \cite{PhysRevLett.104.066403,PhysRevLett.106.197202,PhysRevB.93.161106,PhysRevLett.117.227201,Owerre_2016,Owerre_2017,PhysRevB.90.024412,Li2016,PhysRevB.97.174407,PhysRevB.95.224403,PhysRevB.97.174413,PhysRevLett.117.187203}, acoustic \cite{He2016,PhysRevLett.114.114301}, photonic \cite{RevModPhys.91.015006}, etc..  The possibility of coupling between various degrees of freedom has led to new visions for spintronics \cite{RevModPhys.76.323,ZUTIC201985}, and resulted in new subfields such as spin caloritronics \cite{Bauer2012}, in which spin carriers are manipulated by exciting heat flows. The study of spin currents is fundamental for the field of spintronics, and the spins carried by magnons possess certain advantages over electrons, e.g., low dissipation. At the same time, magnons exhibit rich and fascinating physics associated with the topology of magnonic bands, e.g., the thermal Hall effect has been observed in collinear ferromagent Lu$_2$V$_2$O$_7$ \cite{Onose297}. The spin Nernst effect \cite{Meyer2017,doi:10.1063/1.5021731}, akin to the spin Hall effect \cite{RevModPhys.87.1213}, can also be realized in magnon systems \cite{PhysRevB.93.161106,PhysRevLett.117.217203,PhysRevLett.117.217202,PhysRevB.96.134425,PhysRevB.98.035424,PhysRevB.96.224414,PhysRevB.98.094419}. 

Many spintronics concepts also apply to antiferromagnets \cite{RevModPhys.90.015005}. In particular, collinear antiferromagnets can exhibit the spin Seebeck effect \cite{PhysRevLett.116.097204}, spin pumping \cite{PhysRevLett.113.057601}, spin-orbit torque \cite{RevModPhys.91.035004}, spin Nernst effect \cite{PhysRevLett.117.217203,PhysRevLett.117.217202,PhysRevB.96.134425,PhysRevB.98.035424,PhysRevB.98.094419}, etc. Noncollinear antiferromagnets (NAFMs) have attracted considerable attention in recent years, as such systems support nontrivial band structure topology. The anomalous Hall effect \cite{PhysRevLett.112.017205} and spin Hall effect \cite{PhysRevB.95.075128, PhysRevLett.119.187204} have been realized in Mn$_3$X (X=Ge,Sn,Ga,Ir,Rh, and Pt) systems, where electrons act as charge or spin carriers.  Furthermore, the thermal Hall effect, mediated by magnons, is also identified in NAFM insulators \cite{PhysRevB.98.094419,PhysRevB.99.054409,PhysRevB.95.014422,PhysRevLett.121.097203,PhysRevB.99.014427}. Nevertheless, the magnon-mediated spin transport in NAFMs \cite{PhysRevB.100.100401,PhysRevB.99.224410,2019arXiv190905450M} has not yet been well explored, especially in the context of the topology of magnon bands. As a new class of materials, NAFMs feature rich magnetic point group symmetries, chirality, and easily tunable properties (e.g., by magnetic field). As a result, studies of spin currents in NAFMs can bring new vitality to spintronics, especially in the context of spin caloritronics. In contrast to the unique spin polarization of a magnon current in the collinear system, the spin current in NAFMs can be arbitrarily polarized, which allows a better control of the spin current. NAFMs typically possess different ground states \cite{PhysRevLett.113.237202,PhysRevB.92.144415,PhysRevB.95.094427} depending on the ambient environment, e.g., external field, temperature, substrates, and one can envisage using the spin current as a probe of the ground state.  Meanwhile, many NAFM materials can also hold exotic quantum effects \cite{lacroix2011introduction}. Studies of spin currents in such systems can provide a new venue for probing these materials \cite{Han2019}. Motivated by these interesting possibilities, we initiate a discussion on the magnon-mediated spin current physics in noncollinear antiferromagnets therein and hope to stimulate subsequent research on, e.g., spin transport in topological magnon insulators \cite{PhysRevB.97.081106}, optical generation of magnon-mediated spin currents \cite{PhysRevLett.117.147202,PhysRevLett.122.197702}, and many others, as has been discussed above.

\begin{figure}
	   \includegraphics[width=1\linewidth]{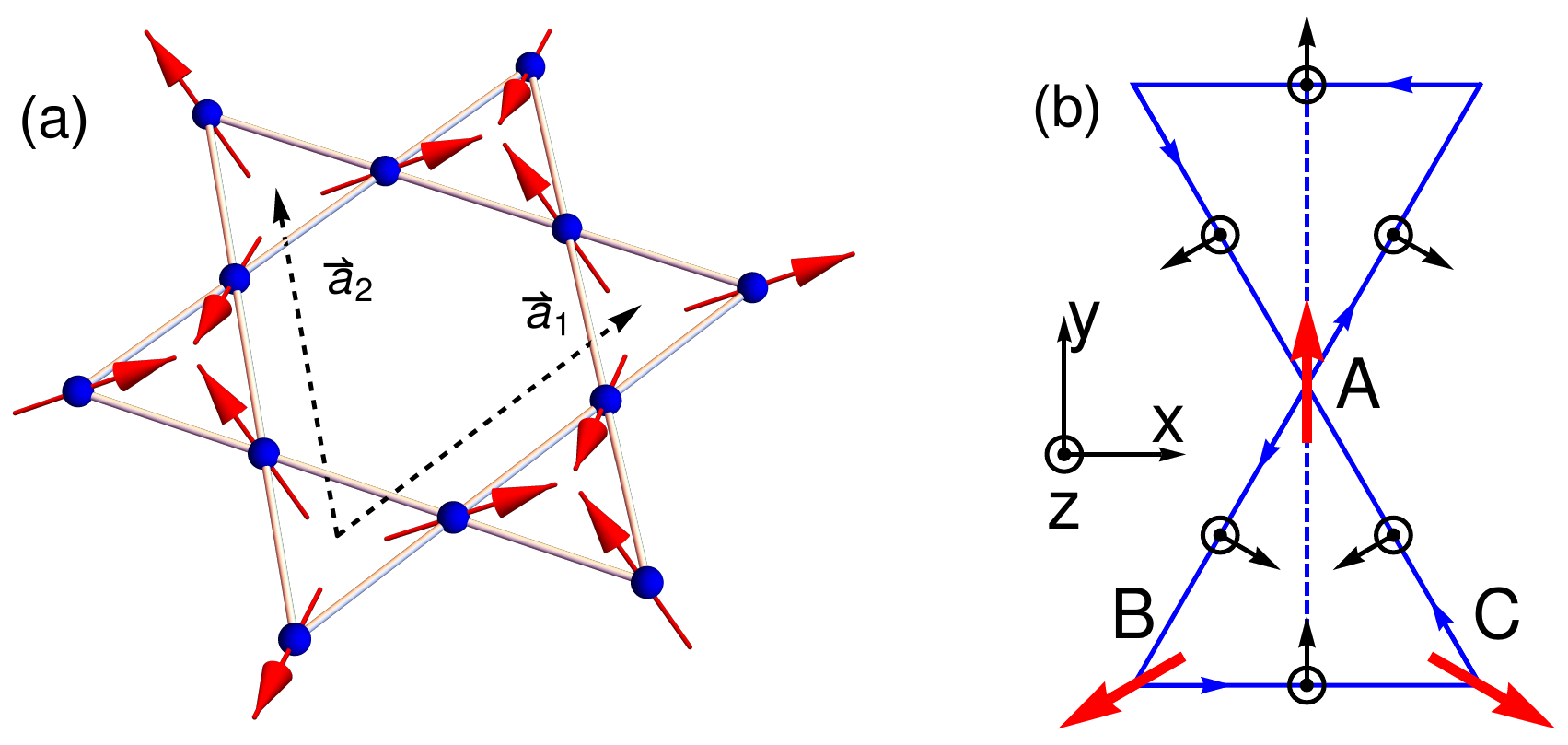}
	\caption{(Color online).(a): Kagome antiferromagnet lattice with small out-of-plane spin canting. (b): Spin order in-plane projection and DMI vectors for kagome antiferromagnet, where the dashed line shows the mirror plane $\mathcal{M}_{x}$.}
	\label{KagomeAF}
\end{figure}

In this paper, we formulate a linear response theory of magnon-mediated spin transport induced by temperature gradients in a noncollinear antiferromagnet, concentrating on the intrinsic contribution not reliant on magnon lifetime.
The difficulty in considering a NAFM is similar to a typical spin Hall system in which spin is not conserved \cite{PhysRevLett.96.076604}. Magnons driven by temperature gradients require accounting for the effects associated with the orbital magnetization~\cite{PhysRevB.89.054420,PhysRevLett.117.217203}.
We start by discussing the definition of spin current in particle-hole space by following Refs.~\cite{PhysRevLett.96.076604,PhysRevLett.117.217203}, where spin non-conservation is signaled by a source term. Next, we develop a linear response theory to temperature gradients for a general observable, i.e., the source term (torque) or spin current, and discuss the symmetry constraints. One of our main results is the expression  for the intrinsic spin Nernst response in noncollinear antiferromagnetic insulators,
\begin{eqnarray}\label{eq:SN}
J_{\gamma\lambda}=\frac{2k_B}{V}\sum_{n=1}^N\sum_{\mathbf{k}}(\Omega_{n,\mathbf{k}}^{j_S})^{\gamma\lambda}_{\beta}c_1[g(\varepsilon_{n,\mathbf{k}})]\nabla_\beta T,
\end{eqnarray}
where $J_{\gamma\lambda}$ is the spin current with polarization $\gamma$, $(\Omega_{n,\mathbf{k}}^{j_S})^{\gamma\lambda}_{\beta}$ is the spin Berry curvature of magnons, and  $c_1(x)=(1+x)\ln(1+x)-x\ln(x)$ is an auxiliary function stemming from the Bose-Einstein statistics of magnons.
We apply our theory to the kagome antiferromagnet KFe$_3$(OH)$_6$(SO$_4$)$_2$ (see Fig.~\ref{KagomeAF}) and show that the in-plane Dzyaloshinskii-Moriya interaction (DMI) leads to a measurable spin Nernst response. Our study opens a way for future studies of fascinating physics related to spin flows in noncollinear antiferromagnets, e.g., in the context of different magnetic orders and material realizations.

\textit{Spin Nernst response}.
We consider a general antiferromagnet with noncollinear ordering. To capture its magnonic excitations at low temperatures, we perform the Holstein-Primakoff transformation in the limit of large $S$, which leads us to a general Hamiltonian  
\begin{equation}\label{eq:ham}
H_0=\frac{1}{2}\int d\mathbf{r}\Psi^\dagger(\mathbf{r})\hat{H}\Psi(\mathbf{r}),
\end{equation}
where $\Psi(\mathbf{r})=(b_1(\mathbf{r})\cdots b_N(\mathbf{r}),b^\dagger(\mathbf{r})\cdots b_N^\dagger(\mathbf{r}))^T$ is the bosonic field, and $N$ the number of atoms in each unit cell. The particle-hole space representation is necessary to describe the anomalous coupling between magnons in an antiferromagnet. 

Due to Bose-Einstein statistics, the eigenvalue problem has to be solved for the matrix $\sigma_3 H_{\mathbf{k}}$     
\cite{doi:10.1063/1.1418246}, where here and in what follows we use Pauli matrices $\sigma_i$ acting in the particle-hole space. Here $H_{\mathbf{k}}$ is the Hamiltonian matrix in momentum space, which can be diagonalized by a paraunitary matrix $T_\mathbf{k}$, i.e., $T_\mathbf{k}^\dagger H_\mathbf{k}T_\mathbf{k}=\mathcal{E}_\mathbf{k}$, where $\mathcal{E}_\mathbf{k}=\text{Diag}(\varepsilon_{1,\mathbf{k}}\cdots \varepsilon_{N,\mathbf{k}},\varepsilon_{1,-\mathbf{k}}\cdots \varepsilon_{N,-\mathbf{k}})$ is the matrix describing eigenvalues, and $T_\mathbf{k}$ satisfies $T^\dagger_\mathbf{k}\sigma_3 T_\mathbf{k}=\sigma_3$. We now write a general theory applicable to bosonic systems where the Bloch wave function corresponding to the band dispersion $\varepsilon_{n,\mathbf{k}}$ is given by $|\psi_{n,\mathbf{k}}\rangle=e^{i \mathbf{k} \cdot \mathbf{r}} |u_{n,\mathbf{k}}\rangle$.  We can then introduce a notation \cite{PhysRevX.7.041045}
\begin{eqnarray}
&&\sigma_3 H_\mathbf{k}|u_{n,\mathbf{k}}^R\rangle=\bar{\varepsilon}_{n,\mathbf{k}}|u_{n,\mathbf{k}}^R\rangle,\nonumber\\
&&\langle u_{n,\mathbf{k}}^L|\sigma_3H_\mathbf{k}=\bar{\varepsilon}_{n,\mathbf{k}}\langle u_{n,\mathbf{k}}^L|,
\end{eqnarray}  
where in terms of the magnonic Hamiltonian $|u_{n,\mathbf{k}}^R\rangle=T_{n,\mathbf{k}}$ and $\langle u^L_{n,\mathbf{k}}|=T_{n,\mathbf{k}}^\dagger\sigma_3$ are the right and left eigenstates of the pseudo-Hermitian Hamiltonian, and $\bar{\varepsilon}_{n,\mathbf{k}}=(\sigma_3\mathcal{E}_\mathbf{k})_{nn}$. Hereafter, we will only refer to the right eigenstates $|u^R_{n,\mathbf{k}}\rangle=|u_{n,\mathbf{k}}\rangle$. The normalization relation reads $\langle u_{n,\mathbf{k}}|\sigma_3|u_{m,\mathbf{k}}\rangle=(\sigma_3)_{nm}$. Moreover, the Hamiltonian (\ref{eq:ham}) possesses particle-hole symmetry (PHS)  so that the Hamiltonian obeys $\sigma_1 H_\mathbf{k}\sigma_1=H_{-\mathbf{k}}^\ast$, which leads to relations $\bar{\varepsilon}_{n+N,\mathbf{k}}=-\bar{\varepsilon}_{n,-\mathbf{k}}$ and $|u_{n,\mathbf{k}}\rangle=e^{i\phi_n}\sigma_1|u_{n+N,-\mathbf{k}}\rangle^\ast$, where $\phi_n$ is a redundant phase factor.

 Because the temperature gradient is a statistical force and doesn't directly enter the Hamiltonian, we introduce a perturbation corresponding to a pseudo-gravitational potential, $\chi(\mathbf{r})$, to account for the temperature gradient
\cite{PhysRev.135.A1505,PhysRevB.89.054420,PhysRevLett.117.217203}, 
\begin{eqnarray}
H^\prime=\frac{1}{4}\int d\mathbf{r}\Psi^\dagger(\mathbf{r})(\chi\hat{H}+\hat{H}\chi)\Psi(\mathbf{r}).
\end{eqnarray}
With the perturbation, the total Hamiltonian is amended to $
H=\frac{1}{2}\int d\mathbf{r}\tilde{\Psi}^\dagger(\mathbf{r})\hat{H}\tilde{\Psi}(\mathbf{r})$,
where $\tilde{\Psi}(\mathbf{r})=(1+\mathbf{r}\cdot\bm{\nabla}\chi/2)\Psi(\mathbf{r})$. To linear order, the system will respond to a temperature gradient in the same way as to a perturbation with $\chi(\mathbf{r})=-T(\mathbf{r})/T$. We now introduce an arbitrary matrix $\hat{O}$ and a local observable $\mathcal{O}(\mathbf{r})=\frac{1}{2}\Psi^\dagger(\mathbf{r})\hat{O}\Psi(\mathbf{r})$. In what follows, we will mostly consider $\hat{O}=\hat{S}^\alpha$, which corresponds to the magnon spin density operator given by  $\hat{S}^\alpha=-\sigma_0\otimes\text{Diag}(\left< S^\alpha_1\right>/S_1, \cdots ,\left< S^\alpha_N\right>/S_N)$, where $\alpha=x,y,z$, $\sigma_0$ describes the particle-hole space, and averages of spins within a unit cell have been taken. The time evolution of this operator can be obtained from the Heisenberg equation applied to the total Hamiltonian (details in Supplemental Material) \cite{PhysRevLett.117.217203}
\begin{eqnarray}
\frac{\partial\mathcal{O}(\mathbf{r})}{\partial t}=i[H,\mathcal{O}(\mathbf{r})]=-\bm{\nabla}\cdot\mathbf{j}_O+S_O.\label{eq:curr}
\end{eqnarray}
Here $\mathbf{j}_O=\tilde{\Psi}^\dagger(\mathbf{r})\hat{\mathbf{j}}_O\tilde{\Psi}(\mathbf{r})$ and $S_O=\tilde{\Psi}^\dagger(\mathbf{r})\hat{S}_O\tilde{\Psi}(\mathbf{r})$ correspond to the local current and source densities, respectively, with $\hat{\mathbf{j}}_O=\frac{1}{4}(\hat{\mathbf{v}}\sigma_3\hat{O}+\hat{O}\sigma_3\hat{\mathbf{v}})$, $\hat{S}_O=-\frac{i}{2}(\hat{O}\sigma_3\hat{H}-\hat{H}\sigma_3\hat{O})$, and $\hat{\mathbf{v}}=i[\hat{H},\mathbf{r}]$. To linear order in the temperature gradient, the above densities are explicitly decomposed as $
\rho_{\theta}=\rho_{\theta}^{[0]}+\rho_{\theta}^{[1]}$, 
with $\rho_{\theta}^{[0]}=\Psi^\dagger(\mathbf{r})\hat{\theta}\Psi(\mathbf{r})$,  $\rho_{\theta}^{[1]}=\frac{1}{2}\Psi^\dagger(\mathbf{r})(\hat{\theta}r_\beta+r_\beta\hat{\theta})\Psi(\mathbf{r})\nabla_\beta\chi$, where for $\theta$ one needs to substitute either $\mathbf{j}_O$ or $S_O$. We will use a four-vector convention in which $\theta_0=S_O$ and $\boldsymbol \theta=\mathbf{j}_O$. 
The non-vanishing source term indicates the non-conservation of the observable, for instance, when $\mathcal{O}(\mathbf{r})$ corresponds to spin density, the source term represents torque density. We note in passing that the source term dipole $\mathbf{P}_O$ can be defined as $S_O=-\bm{\nabla}\cdot \mathbf{P}_O$ for vanishing total source $\frac{1}{V}\int d\mathbf{r} S_O=0$, where $V$ is the volume and $\mathbf{P}_O=\mathbf{r} S_O$. Thus, a conserved current can be defined as $\bm{\mathcal{J}}_O=\mathbf{j}_O+\mathbf{P}_O$ to restore the continuity equation \cite{PhysRevLett.96.076604}. The current term $\mathbf{j}_O$ coincides with the conventional definition in the literature of the spin Hall effect~\cite{RevModPhys.87.1213}. In general, based on Eq.~\eqref{eq:curr} we can interpret $\mathbf{j}_O$ as a spin current and $S_O$ as the torque. In our discussion below, we concentrate on the spin current term. 

We consider spatially averaged quantities $\Theta_\alpha=\Theta_\alpha^{[0]}+\Theta_\alpha^{[1]}$ with $\Theta_\alpha^{[0,1]}=\frac{1}{V}\int d\mathbf{r}\rho^{[0,1]}_{\theta_\alpha}(\mathbf{r})$. The thermal response to linear order in the temperature gradient reads
\begin{eqnarray}\label{eq:response}
\Theta_\alpha=\langle\Theta_\alpha^{[0]}\rangle_{neq}+\langle\Theta_\alpha^{[1]}\rangle_{eq},
\end{eqnarray}
where on the right hand side the first term is evaluated with respect to nonequilibrium states from the Kubo linear response calculation, while the second term corresponds to orbital magnetization in the system and is evaluated with respect to the equilibrium state. In total, we can express the linear response as
$\Theta_\alpha=(S^{\theta_\alpha}_\beta+M^{\theta_\alpha}_\beta)\nabla_\beta\chi$,
where $S^{\theta_\alpha}_\beta$ and $M^{\theta_\alpha}_\beta$ correspond to the first and second terms in Eq.~(\ref{eq:response}). 

In the spirit of the Kubo response calculation \cite{PhysRevB.93.161106, PhysRevLett.117.217203}, the nonequilibrium part can be described by 
 \begin{equation}\label{Kubo}
 \langle \Theta_\alpha^{[0]}\rangle_{\text{neq}}=\lim_{\omega\to 0}\frac{1}{i\omega}[\Pi_{\alpha\beta}(\omega)-\Pi_{\alpha\beta}(0)]\nabla_\beta\chi.
 \end{equation}
 Here $\Pi_{\alpha\gamma}(i\omega_m)=-\int_{0}^{\beta}d\tau e^{i\omega_m\tau}\langle T_\tau \Theta^{[0]}_\alpha(\tau) J_\gamma^q(0)\rangle$, with $\beta=1/(k_B T)$, where $\omega_m$ is the bosonic Matsubara frequency. $\mathbf{J}^q$ is the averaged heat current operator defined as $\mathbf{J}^q=\frac{1}{V}\int d\mathbf{r}\mathbf{j}^q(\mathbf{r})$, where the heat current density $\mathbf{j}^q=\frac{1}{4}\Psi^\dagger(\mathbf{r})(\hat{H}\sigma_3\hat{\mathbf{v}}+\hat{\mathbf{v}}\sigma_3\hat{H})\Psi(\mathbf{r})$. This heat current expression can be inferred from the energy conservation equation $\dot{\rho}_E+\bm{\nabla}\cdot \mathbf{j}^q=0$, where $\rho_E$ is the energy density of the system. After performing the linear response calculation, the intrinsic nonequilibrium coefficient reads 
 \begin{eqnarray}\label{Nonequilibrium}
 S^{\theta_\alpha}_\beta=\frac{1}{V}\sum_{n\mathbf{k}}-\left[(\Omega_{n,\mathbf{k}}^\theta)^{\alpha}_\beta\bar{\varepsilon}_{n,\mathbf{k}} 
 +(m_{n,\mathbf{k}}^\theta)^{\alpha}_\beta \right]g(\bar{\varepsilon}_{n,\mathbf{k}}).
 \end{eqnarray}
 Here
 \begin{eqnarray}
 (\Omega_{n,\mathbf{k}}^\theta)^{\alpha}_\beta=\sum_{m(\neq n)}(\sigma_3)_{nn}(\sigma_3)_{mm}\frac{2\text{Im}[(\theta_{\alpha,\mathbf{k}})_{nm}(v_{\beta,\mathbf{k}})_{mn}]}{(\bar{\varepsilon}_{n,\mathbf{k}}-\bar{\varepsilon}_{m,\mathbf{k}})^2},\nonumber\\
 (m_{n,\mathbf{k}}^\theta)^{\alpha}_\beta=\sum_{m(\neq n)}(\sigma_3)_{nn}(\sigma_3)_{mm}\frac{-\text{Im}[(\theta_{\alpha,\mathbf{k}})_{nm}(v_{\beta,\mathbf{k}})_{mn}]}{\bar{\varepsilon}_{n,\mathbf{k}}-\bar{\varepsilon}_{m,\mathbf{k}}}\nonumber\\
 \end{eqnarray}
 where $(\dots)_{nm}=\langle u_{n,\mathbf{k}}|\dots|u_{m,\mathbf{k}}\rangle$ and $g(\bar{\varepsilon}_{n\mathbf{k}})$ is the Bose-Einstein distribution. Here $(\Omega_{n,\mathbf{k}}^\theta)^{\alpha}_\beta$ is the generalized Berry curvature calculated for operator $\hat{\theta}_\alpha$. This Berry curvature respects the sum rule $\sum_{n=1}^{2N}(\Omega_{n,\mathbf{k}}^\theta)^{\alpha}_\beta=0$, and PHS results in the relation $(\Omega_{n,\mathbf{k}}^\theta)^{\alpha}_\beta=(\Omega_{n+N,-\mathbf{k}}^\theta)^{\alpha}_\beta$ ($1\leq n\leq N$).

 The contribution corresponding to $\rho_{\theta_\alpha}^{[1]}$ is expressed as
\begin{eqnarray}\label{Dipole}
M^{\theta_\alpha}_{\beta}=\frac{1}{2V}\langle\int d\mathbf{r}\Psi^\dagger(\mathbf{r})(\hat{\theta}_\alpha r_\beta+r_\beta\hat{\theta}_\alpha)\Psi(\mathbf{r})\rangle_{eq}.
\end{eqnarray}
To calculate this term, we can identify a  thermodynamic expression for $M^{\theta_\alpha}_\beta$ by following Refs.~\cite{2018arXiv181211721D,PhysRevLett.99.197202,PhysRevB.99.024404,PhysRevB.98.060402,PhysRevB.97.134423}. We introduce a perturbation coupled with a four-component fictitious field $h_\alpha(\mathbf{r})$, i.e., $
\hat{H}_1=\hat{H}_0-[\hat{\theta}_\alpha h_\alpha(\mathbf{r})+h_\alpha(\mathbf{r})\hat{\theta}_\alpha]$. If the field varies very slowly on the scale of the lattice constant, we can identify a  thermodynamic expression
\begin{eqnarray}\label{Dipole1}
M^{\theta_\alpha}_\beta=-\lim_{h_\alpha\rightarrow0}\frac{1}{V}\frac{\partial \Omega}{\partial(\partial_{r_\beta}h_\alpha)}
\end{eqnarray}
where $\Omega$ is the thermodynamic grand potential of the system. If we regard the local fictitious field and its gradient as independent variables, we can assert a Maxwell relation $\left(\partial M^{\theta_\alpha}_\beta/\partial T\right)_{h_\alpha,\partial_{r_\beta} h_\alpha}=\left[\partial S/\partial(\partial_{r_\beta} h_\alpha)\right]_{T,h_\alpha}$, where $S$ is the entropy. Taking both Eq.~\eqref{Dipole1} and the Maxwell relation into account, we are led to 
\begin{eqnarray}\label{Maxwell1}
\tilde{M}^{\theta_\alpha}_{\beta}=\frac{\partial (\beta M^{\theta_\alpha}_{\beta})}{\partial \beta}
\end{eqnarray}
with $\tilde{M}^{\theta_\alpha}_{\beta}=-\frac{1}{V}\frac{\partial K}{\partial(\partial_{r_\beta}h_\alpha)}$ being an auxiliary quantity and $K=\Omega+TS$. We assume that the fictitious field takes the form $h_\alpha(\boldsymbol r)=(h^0_\alpha/q)\sin(\mathbf{q}\cdot \mathbf{r})$, with $\mathbf{q}=q\hat{\mathbf{e}}_\beta$ ($\beta=x,y,z$ in three dimensions and $\beta=x,y$ in two dimensions).
The auxiliary  quantity is calculated by picking up the appropriate Fourier component 
 \begin{eqnarray}\label{Axuiliary}
 \tilde{M}^{\theta_\alpha}_\beta=\lim_{h_\alpha\rightarrow 0}\lim_{q\rightarrow 0}\frac{-2}{h^0_\alpha V}\int d\mathbf{r}\delta K(\mathbf{r})\cos(\mathbf{q}\cdot\mathbf{r})
 \end{eqnarray}
where $\delta K(\mathbf{r})$ is the  variation due to the fictitious field, which can be obtained from perturbation theory \cite{PhysRevLett.99.197202}. Combining Eq.~\eqref{Axuiliary} and \eqref{Maxwell1}, we obtain (see details in Supplemental Material)
\begin{eqnarray}\label{Dipole2}
M^{\theta_\alpha}_\beta=\frac{1}{V}\sum_{n\mathbf{k}}[(\Omega_{n,\mathbf{k}}^\theta)^{\alpha}_\beta\int_0^{\bar{\varepsilon}_{n\mathbf{k}}} d\eta g(\eta) +(m_{n,\mathbf{k}}^\theta)^{\alpha}_\beta g(\bar{\varepsilon}_{n,\mathbf{k}})].\nonumber\\
\end{eqnarray}
By combining the nonequilibrium part in Eq.~\eqref{Nonequilibrium} with Eq.~\eqref{Dipole2} and cancelling the orbital part (corresponding to a bound current), we obtain the thermal response formula which constitutes the main result of this paper:
\begin{eqnarray}\label{Thermalresponse}
\Theta_\alpha=\frac{2k_B}{V}\sum_{n=1}^N\sum_{\mathbf{k}}(\Omega_{n,\mathbf{k}}^\theta)^{\alpha}_{\beta}c_1[g(\varepsilon_{n,\mathbf{k}})]\nabla_\beta T.
\end{eqnarray}
Note that we express our result using particle bands ($n\leq N$) by utilizing PHS.

It is useful to identify the symmetry constraints leading to a vanishing source term response. In general, for the averaged torque density this can happen for only some of the torque components. However, for an inversion symmetric system, i.e., $H_\mathbf{k}=H_{-\mathbf{k}}$, the Berry curvature of the torque term satisfies $(\Omega_{n,\mathbf{k}}^{S_O})_\beta=-(\Omega_{n,-\mathbf{k}}^{S_O})_\beta$. Together with the relation $\varepsilon_{n,\mathbf{k}}=\varepsilon_{n,-\mathbf{k}}$, this results in the vanishing of all torque components in Eq.~\eqref{Thermalresponse}. 

\begin{figure}
    \centering
    \includegraphics[width=1.0\linewidth]{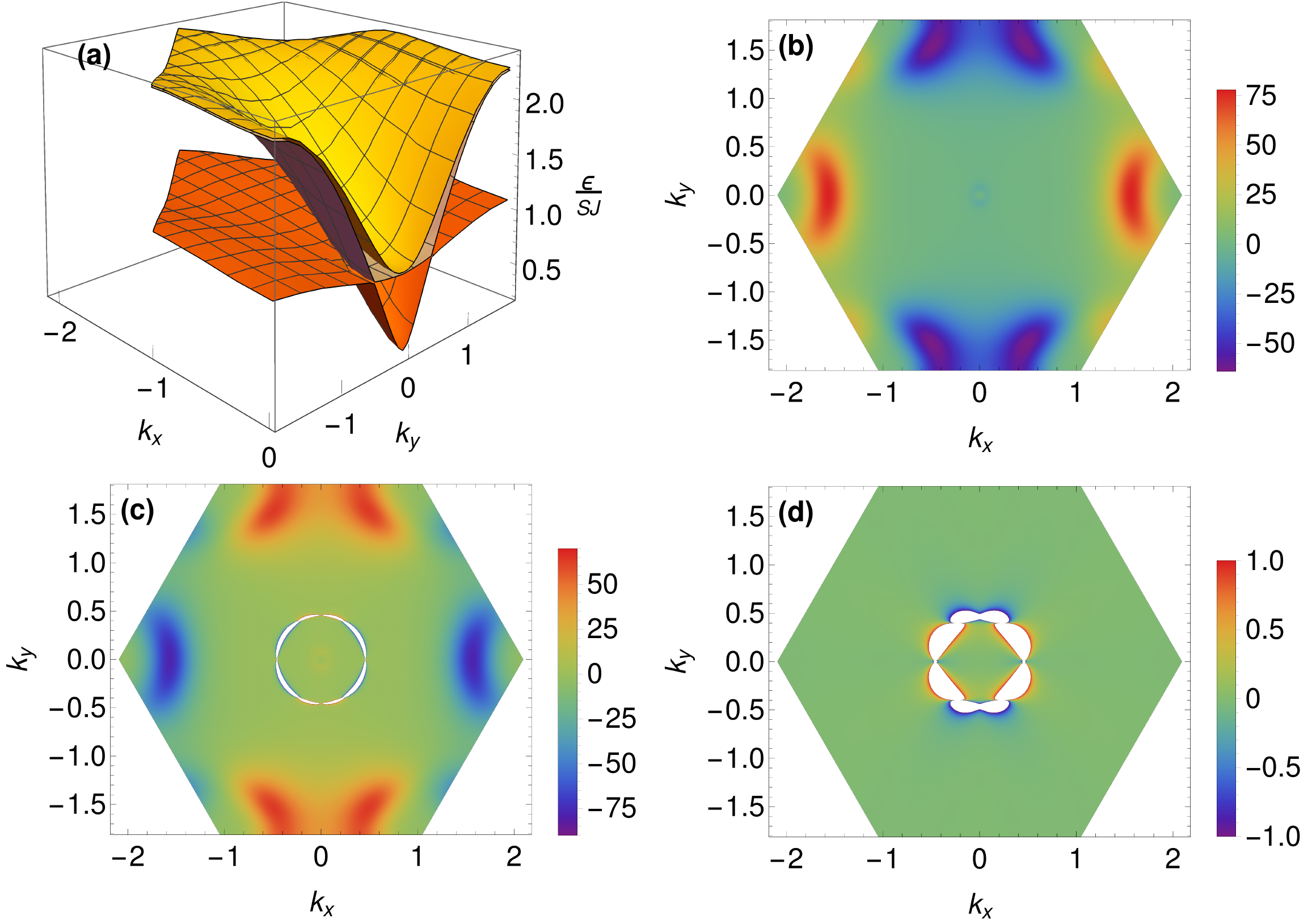}
	\caption{(Color online) Plots for kagome antiferromagnet KFe$_3$(OH)$_6$(SO$_4$)$_2$. (a): Energy bands. (b-d): The spin Berry curvature for $\alpha^y_{yx}$ for top, middle, and lowest band. Detailed plots of the spin Berry curvature in the vicinity of the white regions, corresponding to the values outside of the range of the scale bar, can be found in the Supplemental Material.}
	\label{EnergyandBC}
\end{figure}

\textit{Spin Nernst effect in kagome antiferromagnet}.
We use the result in Eq.~\eqref{Thermalresponse} to calculate the spin Nernst response tensor in a noncollinear kagome antiferromagnet in Eq.~\eqref{eq:SN}
 where the spin Berry curvature is calculated with respect to operator $\hat{j}_{\gamma,\lambda}=\frac{1}{4}(\hat{v}_\lambda\sigma_3\hat{S}^\gamma+\hat{S}^\gamma\sigma_3\hat{v}_\lambda)$ corresponding to the spin current. We can immediately identify that the spin Berry curvature in Eq.~\eqref{eq:SN} is even under the time reversal transformation. As a result, the spin Nernst conductivity is also even under the time reversal transformation, and this result will be used in the symmetry analysis below.  Furthermore, in a kagome antiferromagnet, due to the presence of inversion symmetry, the averaged torque density (source term) vanishes. We consider the Hamiltonian 
\begin{eqnarray}
H=\sum_{\langle ij\rangle}J_1\mathbf{S}_i\cdot\mathbf{S}_j+\mathbf{D}_{ij}\cdot(\mathbf{S}_i\times\mathbf{S}_j)+\sum_{\langle\langle ij\rangle\rangle}J_2 \mathbf{S}_i\cdot\mathbf{S}_j,
\end{eqnarray}
where the first and third terms represent nearest and second-nearest neighbor Heisenberg exchange, and the second term represents nearest neighbor Dzyaloshinskii-Moriya interaction (DMI) with both in-plane and out-of-plane DMI vectors, as shown in Fig.~\ref{KagomeAF}. The DMI vector can be expressed as $\mathbf{D}_{ij}=D_p\hat{n}_{ij}+D_z\hat{z}$, where $D_p$ and $D_z$ correspond to the in-plane and out-of-plane DMI strength, and $\hat{n}_{ij}$ is an in-plane unit vector corresponding to the direction of the in-plane DMI. The in-plane DMI can only arise when $\mathcal{M}_{z}$ symmetry is broken~\cite{PhysRevLett.112.017205}, i.e., time-reversal followed by mirror symmetry with respect to the kagome plane is not a symmetry in such a case. This introduces a small out-of-plane canting angle $\eta$ to spin order with magnitude $\eta=\frac{1}{2}\tan^{-1}(\frac{-2D_p}{\sqrt{3}(J_1+J_2)-D_z})$ \cite{PhysRevB.98.094419}. Here we consider the $\mathbf{q}=0$ phase with spin order as shown in Fig.~\ref{KagomeAF}. The magnetic moments orient according to $\left<\mathbf{S}_{i}\right>=S(\cos\eta\cos\phi_i,\cos\eta\sin\phi_i,\sin\eta)$, where $\phi_i$ is  the angle formed by the in-plane projection of moment with the $x$ axis. Specifically, $\phi_A=\pi/2$, $\phi_B=7\pi/6$, and $\phi_C=-\pi/6$. For the spin Nernst response, we identify $\hat{O}$ discussed above as the spin operator in the magnon basis $\Psi(\mathbf{r})=[b_A(\mathbf{r}),b_B(\mathbf{r}),b_C(\mathbf{r}),b^\dagger_A(\mathbf{r}),b^\dagger_B(\mathbf{r}),b^\dagger_C(\mathbf{r})]^T$, i.e., $\hat{S}^\alpha=-\sigma_0\otimes\text{Diag}(\left<S^\alpha_{A}\right>,\left<S^\alpha_{B}\right>,\left<S^\alpha_{C}\right>)/S$. The spin conductivity tensor of a spin-polarized current in a noncollinear antiferromagnet \cite{PhysRevLett.119.187204,PhysRevB.92.155138} is restricted to a certain form by the magnetic space group of the system. Given that the intrinsic spin Nernst tensor in relation $J_{\gamma\lambda}=\alpha^\gamma_{\lambda\beta}\nabla_\beta T$ is even under the time reversal transformation, the symmetry constraints become
\begin{equation}\label{eq:constr}
    \alpha^\gamma_{\lambda\beta}= \det(R)R_{\gamma\gamma'}R_{\lambda\lambda'}R_{\beta\beta'} \alpha^{\gamma'}_{\lambda'\beta'},
\end{equation}
where the matrix $R$ represents a symmetry element $\mathcal R$ (in Cartesian coordinates) entering the antiunitary symmetry $\mathcal R \mathcal T$ or unitary symmetry $\mathcal R$ of the system (see Supplemental Material). As an example, we focus on a system with two symmetries: mirror reflection with respect to the $y-z$ plane combined with time reversal, $\mathcal{M}_{x}\mathcal{T}$, and threefold rotation about the $z$ axis, $\mathcal{C}_{3z}$. The shape of the spin Nernst tensor corresponding to the constraints in Eq.~\eqref{eq:constr} becomes
\begin{eqnarray}\label{eq:symmetry}
[\hat{\alpha}^{x},\hat{\alpha}^{y},\hat{\alpha}^{z}]=\left[\left(\begin{array}{cc}
-\alpha_1&0\\
0&\alpha_1
\end{array}\right),\right.\left(\begin{array}{cc}0&\alpha_1\\\alpha_1&0
\end{array}\right),\left.\left(\begin{array}{cc}0&-\alpha_2\\\alpha_2&0
\end{array}\right)\right]. 
\end{eqnarray}
Here, the $\mathcal{M}_{x}\mathcal{T}$ symmetry can be replaced by $\mathcal{C}_{2x}\mathcal{T}$, twofold rotation about the $x$ axis and time-reversal, which will lead us to the same result. We note that our results are consistent with the spin Hall response tensors in Mn$_3$X (X= Rh, Ir and Pt)~\cite{PhysRevB.95.075128}.
\begin{figure}
		\includegraphics[width=1\linewidth]{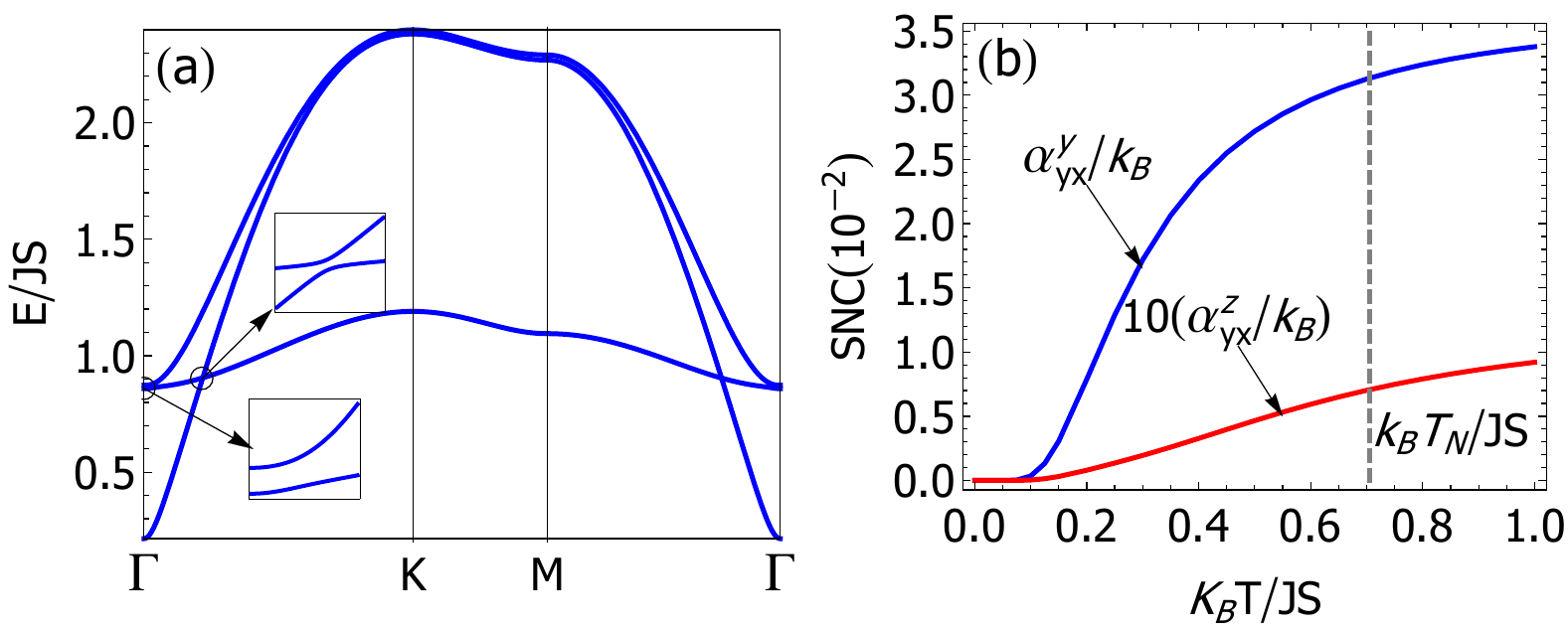}
	\caption{(Color online) Plots for kagome antiferromagnet KFe$_3$(OH)$_6$(SO$_4$)$_2$. (a): Band structure. (b): Spin Nernst conductivity (SNC)  $\alpha^y_{yx}$ and $\alpha^z_{yx}$, where $\alpha^z_{yx}$ is scaled for visibility. Relevant parameters are $J_1=3.18 \text{meV}$, $J_2=0.11\text{meV}$, $|D_p|/J_1=0.062$, $D_z/J_1=-0.062$.}
	\label{SpinNernst_temperature}
\end{figure}

We apply our theory to a single layer of potassium iron jarosite, KFe$_3$(OH)$_6$(SO$_4$)$_2$, for which the material parameters are 
$J_1=3.18 \text{meV}$, $J_2=0.11\text{meV}$, $|D_p|/J_1=0.062$, $D_z/J_1=-0.062$ \cite{PhysRevB.98.094419,PhysRevLett.96.247201}. We note, however, that the magnon dispersion in this material can also be explained by $J_2 = 0$, in which case the flat band is broadened by fluctuations \cite{PhysRevB.92.094409}. The numerically obtained form of the spin Nernst conductivities agrees with Eq.~\eqref{eq:symmetry}. In Fig.~\ref{EnergyandBC}, we plot the magnon bands and the spin Berry curvature for the $y$ polarization of the spin. The spin Berry curvature is peaked at avoided crossings, which give the largest contribution to the spin Nernst effect. The integral of the ordinary Berry curvature gives the Chern numbers $-3$, $1$, and $2$, from the bottom to the top bands in Fig.~\ref{EnergyandBC}. 
In Fig.~\ref{SpinNernst_temperature}, we show the spin Nernst response coefficients as a function of temperature for the $y$ and $z$ spin polarizations. The spin Nernst response sharply increases at temperatures sufficient to excite magnons in the Brillouin zone where the spin Berry curvature is large. The $z$ direction polarized spin current is two orders of magnitude smaller than the current with in-plane spin polarization, which is due to the fact that the canting angle is fairly small, $\eta=1.9^\circ$  \cite{PhysRevB.98.094419}. By applying magnetic field, the canting angle and the spin Nernst response with the $z$ polarization direction can be substantially increased. The predicted spin currents should be easily detectable in three dimensional structures as a temperature gradient of 20 K/mm should result in a spin current of the order of $10^{-11}$ J/m$^2$ according to Fig.~\ref{SpinNernst_temperature}, where $\alpha^{3D}=\alpha/c$, with $c$ being the interlayer distance. Finally, we note that the spin Nernst effect reported in Ref.~\cite{PhysRevB.100.100401} differs from the intrinsic effect reported here as the former has the symmetry of the extrinsic effect.

\textit{Conclusions.} 
We have developed a theory of magnon-mediated intrinsic spin currents in insulating noncollinear antiferromagnets and applied this theory to  potassium iron jarosite KFe$_3$(OH)$_6$(SO$_4$)$_2$. Our results are applicable to two- and three-dimensional systems, promising to reveal fascinating physics in other layered quasi-2D antiferromagnets, e.g., silver iron jarosite AgFe$_3$(OH)$_6$(SO$4$)$_2$ \cite{PhysRevB.83.214406}, chromium jarosite KCr$_3$(OH)$_6$(SO$4$)$_2$ \cite{PhysRevB.64.054421}, vesignieite BaCu$_3$V$_2$O$_8$(OH)$_2$ \cite{PhysRevB.88.144419}, and 3D pyrochlore antiferromagnets LiGaGr$_4$O$_8$ and LiInGr$_4$O$_8$ \cite{PhysRevB.90.060414,PhysRevLett.113.227204}. 
Besides exploring material candidates, one can also study the effect of magnetic order on the spin Nernst effect, e.g., in kagome antiferromagnets other possible spin chiralities exist \cite{PhysRevB.95.094427,PhysRevB.99.014427}. Recently proposed antiferromagnetic skyrmions with noncollinear magnetic order \cite{PhysRevLett.122.187203} can also be explored using our theory. 

We gratefully acknowledge stimulating discussions
with Kirill Belashchenko, Ding-Fu Shao and Liang Dong. This work was supported by the U.S. Department of Energy, Office of Science, Basic Energy
Sciences, under Award No. DE-SC0014189.

\bibliographystyle{apsrev}
\bibliography{SpinNernst}

\newpage
\begin{widetext}
\begin{center}
	{{\bf Supplemental material}}
\end{center}

\section{Time evolution of a local observable}
To derive the time evolution equation for a local observable $\mathcal{O}(\mathbf{r})=\frac{1}{2}\Psi^\dagger(\mathbf{r})\hat{O}\Psi(\mathbf{r})$, we first prepare a basic knowledge of the Hamiltonian operator and commutators in particle-hole space by following Ref.~\cite{PhysRevLett.117.217203,PhysRevB.89.054420}. The total Hamiltonian can be generally expressed as $H=\frac{1}{2}\int d\mathbf{r}\tilde{\Psi}^\dagger(\mathbf{r})\hat{H}\tilde{\Psi}(\mathbf{r})$ with $\hat{H}=\sum_{\bm{\delta}}H_{\bm{\delta}}e^{i\hat{\mathbf{p}}\cdot\bm{\delta}}$, in which $e^{i\hat{\mathbf{p}}\cdot\bm{\delta}}$ is the translation operator that satisfies $e^{i\hat{\mathbf{p}}\cdot\bm{\delta}}f(\mathbf{r})=f(\mathbf{r}+\bm{\delta})$. Here $\bm{\delta}$ is the vector shift between unit cells, $\tilde{\Psi}(\mathbf{r})=(1+\mathbf{r}\cdot\bm{\nabla}\chi/2)\Psi(\mathbf{r})$. Based on the basic commutators between bosons $[a_i(\mathbf{r}),a^\dagger_j(\mathbf{r^\prime})]=\delta_{ij}\delta_{\mathbf{r},\mathbf{r}^\prime}$, $[a_i(\mathbf{r}),a_j(\mathbf{r^\prime})]=0$, we can construct commutators in the particle-hole basis
\begin{eqnarray}
[\Psi_i(\mathbf{r}),\Psi_j^\dagger(\mathbf{r}^\prime)]=(\sigma_3)_{ij}\delta_{\mathbf{r},\mathbf{r^\prime}},\qquad
[\Psi_i(\mathbf{r}),\Psi_j(\mathbf{r}^\prime)]=i(\sigma_2)_{ij}\delta_{\mathbf{r},\mathbf{r^\prime}},\qquad
[\Psi^\dagger_i(\mathbf{r}),\Psi_j^\dagger(\mathbf{r}^\prime)]=-i(\sigma_2)_{ij}\delta_{\mathbf{r},\mathbf{r^\prime}}
\end{eqnarray}
where $\sigma_{i}$ ($i=1,2,3$) are Pauli matrices acting in particle-hole space. Now we use the above Hamiltonian and commutators to perform a local observable time evolution calculation in two steps. First, we work out the Heisenberg equation commutation as follow,
\begin{eqnarray}
\frac{\partial\mathcal{O}(\mathbf{r})}{\partial t}=&&i[H,\mathcal{O}(\mathbf{r})]=i[\frac{1}{2}\sum_{\bm{\delta}}\int d\mathbf{r}^\prime \tilde{\Psi}^\dagger(\mathbf{r}^\prime)H_{\bm{\delta}}\tilde{\Psi}(\mathbf{r}^\prime+\bm{\delta}),\frac{1}{2}\Psi^\dagger(\mathbf{r})\hat{O}\Psi(\mathbf{r})]\nonumber\\
&&=\frac{i}{4}\sum_{\bm{\delta}}\int d\mathbf{r}^\prime\xi(\mathbf{r}^\prime)(H_{\bm{\delta}})_{ij}\xi(\mathbf{r}^\prime+\bm{\delta})O_{mn}[\Psi^\dagger_i(\mathbf{r}^\prime)\Psi_j(\mathbf{r}^\prime+\bm{\delta}),\Psi^\dagger_m(\mathbf{r})\Psi_n(\mathbf{r})]\nonumber\\
=&&-\frac{i}{2}\sum_{\bm{\delta}}[\tilde{\Psi}^\dagger(\mathbf{r})\hat{O}\sigma_3H_{\bm{\delta}}\tilde{\Psi}(\mathbf{r}+\bm{\delta})-
\tilde{\Psi}^\dagger(\mathbf{r}-\bm{\delta})H_{\bm{\delta}}\sigma_3\hat{O}\tilde{\Psi}(\mathbf{r})].
\end{eqnarray}
Here we used the simplified notation $\xi(\mathbf{r})=1+\mathbf{r}\cdot\bm{\nabla}\chi/2$. We also took advantage of particle-hole symmetry, i.e., $\Psi_n(\mathbf{r})=(\sigma_1)_{nl}\Psi^\dagger_l(\mathbf{r})$ and $\sigma_1\hat{O}\sigma_1=\hat{O}$, where the second relation results from the first one. Next, we reduce the above result to a continuous expression by properly sending the shift vector to an infinitely small value.
\begin{eqnarray}
\frac{\partial\mathcal{O}(\mathbf{r})}{\partial t}=&&-\frac{i}{2}\sum_{\bm{\delta}}[\tilde{\Psi}^\dagger(\mathbf{r})\hat{O}\sigma_3H_{\bm{\delta}}\tilde{\Psi}(\mathbf{r}+\bm{\delta})-
\tilde{\Psi}^\dagger(\mathbf{r}-\bm{\delta})H_{\bm{\delta}}\sigma_3\hat{O}\tilde{\Psi}(\mathbf{r})]\nonumber\\
=&&-\frac{1}{2}\sum_{\bm{\delta}}\frac{1}{\bm{\delta}}[\tilde{\Psi}^\dagger(\mathbf{r})\hat{O}\sigma_3 (i\bm{\delta}H_{\bm{\delta}}e^{i\hat{\mathbf{p}}\cdot\bm{\delta}})\tilde{\Psi}(\mathbf{r})-\tilde{\Psi}^\dagger(\mathbf{r}-\bm{\delta}) (i\bm{\delta}H_{\bm{\delta}}e^{i\hat{\mathbf{p}}\cdot\bm{\delta}})\sigma_3\hat{O}\tilde{\Psi}(\mathbf{r}-\bm{\delta})]\nonumber
\\
=&&-\frac{1}{2}\sum_{\bm{\delta}}\frac{1}{\bm{\delta}}[\tilde{\Psi}^\dagger(\mathbf{r})\frac{1}{2}(\hat{O}\sigma_3\hat{v}_{\bm{\delta}}+\hat{v}_{\bm{\delta}}\sigma_3\hat{O})\tilde{\Psi}(\mathbf{r})+\tilde{\Psi}^\dagger(\mathbf{r})\frac{1}{2}(\hat{O}\sigma_3\hat{v}_{\bm{\delta}}-\hat{v}_{\bm{\delta}}\sigma_3\hat{O})\tilde{\Psi}(\mathbf{r})]-\frac{1}{\bm{\delta}}[\tilde{\Psi}^\dagger(\mathbf{r}-\bm{\delta})\frac{1}{2}(\hat{O}\sigma_3\hat{v}_{\bm{\delta}}+\hat{v}_{\bm{\delta}}\sigma_3\hat{O})\tilde{\Psi}(\mathbf{r}-\bm{\delta})\nonumber\\
&&-\tilde{\Psi}^\dagger(\mathbf{r}-\bm{\delta})\frac{1}{2}(\hat{O}\sigma_3\hat{v}_{\bm{\delta}}-\hat{v}_{\bm{\delta}}\sigma_3\hat{O})\tilde{\Psi}(\mathbf{r}-\bm{\delta})]\nonumber
\\
=&&-\frac{1}{4}\sum_{\bm{\delta}}\frac{1}{\bm{\delta}}[\tilde{\Psi}^\dagger(\mathbf{r})(\hat{O}\sigma_3\hat{v}_{\bm{\delta}}+\hat{v}_{\bm{\delta}}\sigma_3\hat{O})\tilde{\Psi}(\mathbf{r})-\tilde{\Psi}^\dagger(\mathbf{r}-\bm{\delta})(\hat{O}\sigma_3\hat{v}_{\bm{\delta}}+\hat{v}_{\bm{\delta}}\sigma_3\hat{O})\tilde{\Psi}(\mathbf{r}-\bm{\delta})]-\frac{1}{4}\sum_{\bm{\delta}}\frac{1}{\bm{\delta}}[\tilde{\Psi}^\dagger(\mathbf{r})(\hat{O}\sigma_3\hat{v}_{\bm{\delta}}-\hat{v}_{\bm{\delta}}\sigma_3\hat{O})\tilde{\Psi}(\mathbf{r})\nonumber\\
&&+\tilde{\Psi}^\dagger(\mathbf{r}-\bm{\delta})(\hat{O}\sigma_3\hat{v}_{\bm{\delta}}-\hat{v}_{\bm{\delta}}\sigma_3\hat{O})\tilde{\Psi}(\mathbf{r}-\bm{\delta})]\nonumber\\
&&=-\frac{1}{4}\bm{\nabla}\cdot[\tilde{\Psi}^\dagger(\mathbf{r})(\hat{O}\sigma_3\hat{\mathbf{v}}+\hat{\mathbf{v}}\sigma_3\hat{O})\tilde{\Psi}(\mathbf{r})]-\frac{i}{2}
\tilde{\Psi}^\dagger(\mathbf{r})(\hat{O}\sigma_3\hat{H}-\hat{H}\sigma_3\hat{O})\tilde{\Psi}(\mathbf{r}).
\end{eqnarray}
Here we used the notation $\hat{v}_{\bm{\delta}}=i\bm{\delta}H_{\bm{\delta}}e^{i\hat{\mathbf{p}}\cdot\bm{\delta}}$ and $\hat{\mathbf{v}}=i\sum_{\bm{\delta}}\bm{\delta}H_{\bm{\delta}}e^{i\hat{\mathbf{p}}\cdot\bm{\delta}}=i[\hat{H},\mathbf{r}]$.
In the last line, we take the limit $\bm{\delta}\rightarrow 0$ to obtain the continuous expression. We can easily read out the current and source term as discussed in the main text from the final result.

\section{Linear response theory}
We provide a fully quantum mechanical derivation in this section. As shown in the main text, the linear thermal response for a given observable can be expressed as
\begin{eqnarray}
\Theta_\alpha=L^{\theta_\alpha}_\beta\nabla_\beta\chi=(S^{\theta_\alpha}_\beta+M^{\theta_\alpha}_\beta)\nabla_\beta\chi,
\end{eqnarray}
where $S^{\theta_\alpha}_\beta$ and $M^{\theta_\alpha}_\beta$ are the Kubo response and dipole moment contribution, respectively. In the following, we will first calculate the dipole moment part from a thermodynamic point of view and then combine it with the Kubo part to arrive at the final response formula. All calculations will be performed in the full particle-hole space, and we will express the final result in terms of particle space in the end.

\subsection{Dipole moment contribution}
In the main text, we have shown the relation $M^{\theta_\alpha}_\beta=-\lim_{h_\alpha\rightarrow0}\frac{1}{V}\frac{\partial \Omega}{\partial(\partial_{r_\beta}h_\alpha)}$. The thermodynamic definition of grand potential reads  $\Omega=E-TS-\mu N$, where $S$, $\mu$, $N$ are the entropy, chemical potential, particle number of the system, and $E$ is the energy which reads
\begin{eqnarray}
E=\langle H\rangle_{eq}=\frac{1}{2}\sum_{\mathbf{k},n=1}^{2N}(\sigma_3)_{nn}g(\bar{\varepsilon}_{n,\mathbf{k}})\langle\psi_{n,\mathbf{k}}|\hat{H}|\psi_{n,\mathbf{k}}\rangle.
\end{eqnarray}
Here we use the relation $\langle\Gamma^\dagger_{n,\mathbf{k}}\Gamma_{m,\mathbf{k}}\rangle=(\sigma_3)_{nn}g(\bar{\varepsilon}_{n,\mathbf{k}})$ with $\Gamma_{m,\mathbf{k}}=\sum_l(T_{\mathbf{k}})_{ml}\Psi_{\mathbf{k},l}$. Below we will assume the chemical potential to be zero. If we regard the local fictitious field and its gradient as independent variables, the variation of the grand potential can be identified as 
\begin{eqnarray}
d\Omega=-SdT-\langle\Theta^{[0]}_\alpha\rangle dh_\alpha-M^{\theta_\alpha}_\beta d(\partial_{r_\beta}h_\alpha),
\end{eqnarray}
from which we can identify the Maxwell relation
\begin{eqnarray}\label{Maxwell}
\left(\frac{\partial M^{\theta_\alpha}_\beta}{\partial T}\right)_{h_\alpha,\partial_{r_\beta} h_\alpha}=\left[\frac{\partial S}{\partial(\partial_{r_\beta} h_\alpha)}\right]_{T,h_\alpha}.
\end{eqnarray}
To get rid of calculations involving entropy $S$, we first introduce an auxiliary quantity 
\begin{eqnarray}
\tilde{M}^{\theta_\alpha}_{\beta}=-\frac{1}{V}\frac{\partial K}{\partial(\partial_{r_\beta}h_\alpha)}
\end{eqnarray}
with $K=\Omega+TS=E$ ($\mu=0$).
By utilizing Eq.~\eqref{Maxwell}, we obtain
\begin{eqnarray}
M^{\theta_\alpha}_\beta=\tilde{M}^{\theta_\alpha}_\beta+T\frac{\partial M^{\theta_\alpha}_\beta}{\partial T}
\end{eqnarray}
and hence the dipole moment contribution can be calculated as 
\begin{eqnarray}\label{Dipolerelation}
\tilde{M}^{\theta_\alpha}_{\beta}=\frac{\partial (\beta M^{\theta_\alpha}_{\beta})}{\partial \beta}.
\end{eqnarray}
If we regard the fictitious field term as a perturbation, the variation of $K$ to linear order reads
\begin{eqnarray}
\delta K(\mathbf{r})&&=\frac{1}{2}\sum_{n\mathbf{k}}\delta g(\bar{\varepsilon}_{n\mathbf{k}})(\sigma_3)_{nn}\langle\psi_{n\mathbf{k}}|\hat{H}|\psi_{n\mathbf{k}}\rangle-g(\bar{\varepsilon}_{n\mathbf{k}})(\sigma_3)_{nn}\langle\psi_{n\mathbf{k}}|[\hat{\theta}_\alpha h_\alpha(\mathbf{r})+h_\alpha(\mathbf{r})\hat{\theta}_\alpha]|\psi_{n\mathbf{k}}\rangle\nonumber\\
&&+g(\bar{\varepsilon}_{n\mathbf{k}})(\sigma_3)_{nn}(\langle\delta\psi_{n\mathbf{k}}|\hat{H}|\psi_{n\mathbf{k}}\rangle+\langle\psi_{n\mathbf{k}}|\hat{H}|\delta\psi_{n\mathbf{k}}\rangle),
\end{eqnarray}
where $|\psi_{n\mathbf{k}}\rangle=\frac{e^{i\mathbf{k}\cdot\mathbf{r}}}{\sqrt{V}}|u_{n\mathbf{k}}\rangle$ is the Bloch wave function of the system.  If we assume a special form of the fictitious field
\begin{eqnarray}
h_\alpha(\mathbf{r})=\frac{h^0_\alpha}{q}\sin(\mathbf{q}\cdot\mathbf{r}),
\end{eqnarray}
with $\mathbf{q}=q\hat{\mathbf{e}}_\beta$, where $\alpha,\beta=x,y,z$ in three dimensions or $\alpha,\beta=x,y$ in two dimensions, the auxiliary quantity can be identified by picking up the appropriate Fourier component
\begin{eqnarray}
\tilde{M}^{\theta_\alpha}_\beta=\lim_{q\rightarrow 0}\frac{-2}{h^0_\alpha V}\int d\mathbf{r}\delta K(\mathbf{r})\cos(\mathbf{q}\cdot\mathbf{r}).
\end{eqnarray}
As an example, we calculate $\tilde{M}^{\theta_y}_x$ by taking $\mathbf{q}_1=q\hat{\mathbf{e}}_x$ and $h_\alpha(\mathbf{r})=\frac{h}{q}\sin(\mathbf{q}_1\cdot\mathbf{r})\delta_{\alpha,y}$. Applying perturbation theory to linear order under the Bloch representation, we find
\begin{eqnarray}
\langle\psi_{m,\mathbf{k}\pm\mathbf{q}_1}|\sigma_3|\delta\psi_{n\mathbf{k}}\rangle=\frac{ih}{2q}\frac{\langle u_{m,\mathbf{k}\pm\mathbf{q}_1}|(\theta_{y,\mathbf{k}}+\theta_{y,\mathbf{k}+\mathbf{q}_1})|u_{n,\mathbf{k}}\rangle}{\bar{\varepsilon}_{n\mathbf{k}}-\bar{\varepsilon}_{m,\mathbf{k}\pm\mathbf{q}_1}},
\end{eqnarray}
and
\begin{eqnarray}
|\delta\psi_{n\mathbf{k}}\rangle=\sum_{m}\frac{ih}{2q}(\sigma_3)_{mm}[e^{i(\mathbf{k}+\mathbf{q}_1)\cdot\mathbf{r}}|u_{m,\mathbf{k}+\mathbf{q}_1}\rangle \frac{\langle u_{m,\mathbf{k}+\mathbf{q}_1}|(\theta_{y,\mathbf{k}}+\theta_{y,\mathbf{k}+\mathbf{q}_1})|u_{n,\mathbf{k}}\rangle}{\bar{\varepsilon}_{n\mathbf{k}}-\bar{\varepsilon}_{m,\mathbf{k}+\mathbf{q}_1}}-(\mathbf{q}_1\rightarrow-\mathbf{q}_1)].
\end{eqnarray}
Here it is implied that we will use the operator under Bloch representation henceforth, i.e., $\hat{H}\rightarrow H_\mathbf{k}=e^{-i\mathbf{k}\cdot\mathbf{r}}\hat{H}e^{i\mathbf{k}\cdot\mathbf{r}}$, $\hat{\theta}_\alpha\rightarrow \theta_{\alpha,\mathbf{k}}=e^{-i\mathbf{k}\cdot\mathbf{r}}\hat{\theta}_\alpha e^{i\mathbf{k}\cdot\mathbf{r}}$. This step is guaranteed by the requirement that the operator $\hat{\theta}_\alpha$ is well defined in a periodic system. By using the results above we obtain 
\begin{eqnarray}
\tilde{M}^{\theta_y}_x&&=\lim_{q\rightarrow0}\frac{1}{2V}\sum_\mathbf{k}\sum_{mn}\frac{1}{i2q}g(\bar{\varepsilon}_{n\mathbf{k}})(\sigma_3)_{nn}(\sigma_3)_{mm}\bar{\varepsilon}_{n\mathbf{k}}[\frac{\langle u_{n\mathbf{k}}|\sigma_3| u_{m,\mathbf{k}+\mathbf{q}_1}\rangle\langle u_{m,\mathbf{k}+\mathbf{q}_1}|(\theta_{y,\mathbf{k}}+\theta_{y,\mathbf{k}+\mathbf{q}_1})|u_{n,\mathbf{k}}\rangle}{\bar{\varepsilon}_{n\mathbf{k}}-\bar{\varepsilon}_{m,\mathbf{k}+\mathbf{q}_1}}-(\mathbf{q}_1\rightarrow-\mathbf{q}_1)]+c.c.\nonumber\\
&&=\lim_{q\rightarrow0}\frac{1}{2V}\sum_\mathbf{k}\sum_{mn}\frac{1}{i2q}[g(\bar{\varepsilon}_{n\mathbf{k}})\bar{\varepsilon}_{n\mathbf{k}}-g(\bar{\varepsilon}_{m,\mathbf{k}+\mathbf{q}_1})\bar{\varepsilon}_{m,\mathbf{k}+\mathbf{q}_1}](\sigma_3)_{nn}(\sigma_3)_{mm}\frac{\langle u_{n\mathbf{k}}|\sigma_3| u_{m,\mathbf{k}+\mathbf{q}_1}\rangle\langle u_{m,\mathbf{k}+\mathbf{q}_1}|(\theta_{y,\mathbf{k}}+\theta_{y,\mathbf{k}+\mathbf{q}_1})|u_{n,\mathbf{k}}\rangle}{\bar{\varepsilon}_{n\mathbf{k}}-\bar{\varepsilon}_{m,\mathbf{k}+\mathbf{q}_1}}+c.c.\nonumber\\
\end{eqnarray}
Taking the limit, we get for $m\neq n$,
\begin{eqnarray}
(\tilde{M}^{\theta_y}_x)_1&&=\frac{1}{V}\sum_\mathbf{k}\sum_{m\neq n}\frac{1}{2}[g(\bar{\varepsilon}_{m\mathbf{k}})\bar{\varepsilon}_{m\mathbf{k}}-g(\bar{\varepsilon}_{n,\mathbf{k}})\bar{\varepsilon}_{n,\mathbf{k}}](\sigma_3)_{nn}(\sigma_3)_{mm}\frac{i\langle u_{n,\mathbf{k}}|\sigma_3|\partial_{k_x} u_{m,\mathbf{k}}\rangle\langle u_{m,\mathbf{k}}|\theta_y|u_{n,\mathbf{k}}\rangle}{\bar{\varepsilon}_{n,\mathbf{k}}-\bar{\varepsilon}_{m,\mathbf{k}}}+c.c.\nonumber\\
&&=\frac{1}{V}\sum_\mathbf{k}\sum_{m\neq n}-\frac{1}{2}[g(\bar{\varepsilon}_{m\mathbf{k}})\bar{\varepsilon}_{m\mathbf{k}}-g(\bar{\varepsilon}_{n,\mathbf{k}})\bar{\varepsilon}_{n,\mathbf{k}}](\sigma_3)_{nn}(\sigma_3)_{mm}\frac{i\langle u_{n,\mathbf{k}}|v_x| u_{m,\mathbf{k}}\rangle\langle u_{m,\mathbf{k}}|\theta_y|u_{n,\mathbf{k}}\rangle}{(\bar{\varepsilon}_{n,\mathbf{k}}-\bar{\varepsilon}_{m,\mathbf{k}})^2}+c.c..
\end{eqnarray}
For $m=n$, we have
\begin{eqnarray}
(\tilde{M}^{\theta_y}_x)_2&&=\frac{1}{V}\sum_\mathbf{k}\sum_n\frac{1}{2i}[g(\bar{\varepsilon}_{n,\mathbf{k}})+g^\prime(\bar{\varepsilon}_{n,\mathbf{k}})\bar{\varepsilon}_{n,\mathbf{k}}][\langle u_{n,\mathbf{k}}|\sigma_3\partial_{k_x}u_{n,\mathbf{k}}\rangle\langle u_{n,\mathbf{k}}|\theta_y|u_{n,\mathbf{k}}\rangle+(\sigma_3)_{nn}\langle\partial_{k_x}u_{n,\mathbf{k}}|\theta_y|u_{n,\mathbf{k}}\rangle]+c.c.\nonumber\\
&&=\frac{1}{V}\sum_\mathbf{k}\sum_n-\frac{1}{2}[g(\bar{\varepsilon}_{n,\mathbf{k}})+g^\prime(\bar{\varepsilon}_{n,\mathbf{k}})\bar{\varepsilon}_{n,\mathbf{k}}](\sigma_3)_{nn}(\sigma_3)_{mm}\frac{i\langle u_{n,\mathbf{k}}|v_x| u_{m,\mathbf{k}}\rangle\langle u_{m,\mathbf{k}}|\theta_y|u_{n,\mathbf{k}}\rangle}{\bar{\varepsilon}_{n,\mathbf{k}}-\bar{\varepsilon}_{m,\mathbf{k}}}+c.c..
\end{eqnarray}
Above $v_x=\partial_{k_x}H$.   In total, we have
\begin{eqnarray}
\tilde{M}^{\theta_y}_x&&=(\tilde{M}^{\theta_y}_x)_1+(\tilde{M}^{\theta_y}_x)_2=\frac{1}{V}\sum_{n\mathbf{k}}g(\bar{\varepsilon}_{n\mathbf{k}})\bar{\varepsilon}_{n\mathbf{k}}(\Omega^\theta_{n,\mathbf{k}})^y_x+[g(\bar{\varepsilon}_{n,\mathbf{k}})+g^\prime(\bar{\varepsilon}_{n,\mathbf{k}})\bar{\varepsilon}_{n,\mathbf{k}}](m_{n,\mathbf{k}}^\theta)^y_x.
\end{eqnarray}
The calculation of all other components is fully analogous to what we have done. The general result will be 
\begin{eqnarray}
\tilde{M}^{\theta_\alpha}_\beta=\frac{1}{V}\sum_{n\mathbf{k}}g(\bar{\varepsilon}_{n\mathbf{k}})\bar{\varepsilon}_{n\mathbf{k}}(\Omega^\theta_{n,\mathbf{k}})^{\alpha}_\beta+[g(\bar{\varepsilon}_{n,\mathbf{k}})+g^\prime(\bar{\varepsilon}_{n,\mathbf{k}})\bar{\varepsilon}_{n,\mathbf{k}}](m^\theta_{n,\mathbf{k}})^{\alpha}_\beta,
\end{eqnarray}
with
\begin{eqnarray}
(\Omega^\theta_{n,\mathbf{k}})^{\alpha}_\beta&&=\sum_{m(\neq n)}(\sigma_3)_{nn}(\sigma_3)_{mm}\frac{2\text{Im}(\langle u_{n,\mathbf{k}}|\theta_\alpha| u_{m,\mathbf{k}}\rangle\langle u_{m,\mathbf{k}}|v_\beta|u_{n,\mathbf{k}}\rangle)}{(\bar{\varepsilon}_{n,\mathbf{k}}-\bar{\varepsilon}_{m,\mathbf{k}})^2},\label{Berrycurvature1}\\
(m^\theta_{n,\mathbf{k}})^{\alpha}_\beta&&=-\sum_{m(\neq n)}(\sigma_3)_{nn}(\sigma_3)_{mm}\frac{\text{Im}(\langle u_{n,\mathbf{k}}|\theta_\alpha| u_{m,\mathbf{k}}\rangle\langle u_{m,\mathbf{k}}|v_\beta|u_{n,\mathbf{k}}\rangle)}{\bar{\varepsilon}_{n,\mathbf{k}}-\bar{\varepsilon}_{m,\mathbf{k}}}.
\end{eqnarray}
Note the Berry curvature defined in Eq.~\eqref{Berrycurvature1} exists in both particle and hole space.
Finally, by using Eq.~\eqref{Dipolerelation} we obtain 
\begin{eqnarray}
M^{\theta_\alpha}_\beta=\frac{1}{\beta}\int_0^\beta d\bar{\beta}\tilde{M}^{\theta_\alpha}_\beta=\frac{1}{V}\sum_{n\mathbf{k}}[(\Omega^\theta_{n,\mathbf{k}})^{\alpha}_\beta\int_0^{\bar{\varepsilon}_{n\mathbf{k}}} d\eta g(\eta) +(m^\theta_{n,\mathbf{k}})^{\alpha}_\beta g(\bar{\varepsilon}_{n,\mathbf{k}})].
\end{eqnarray}
Here we used the relation $\frac{1}{\beta}\int_0^\beta d\bar{\beta} g(\bar{\varepsilon}_{n,\mathbf{k}})\bar{\varepsilon}_{n,\mathbf{k}}=\int_0^{\bar{\varepsilon}_{n\mathbf{k}}}d\eta g(\eta)$ with $g(\eta)=\frac{1}{e^{\bar{\beta}\eta}-1}$ and $\frac{d}{d\bar{\beta}}[\bar{\beta}g(\bar{\varepsilon}_{n,\mathbf{k}})]=g(\bar{\varepsilon}_{n,\mathbf{k}})+g^\prime(\bar{\varepsilon}_{n,\mathbf{k}})\bar{\varepsilon}_{n,\mathbf{k}}$.

\subsection{Kubo type response}
The intrinsic part of the Kubo linear response in particle-hole space is
\begin{eqnarray}
S^{\theta_\alpha}_\beta&&=\frac{1}{V}\sum_{m\neq n}\sum_\mathbf{k}\frac{i}{2}(\theta_{\alpha,\mathbf{k}})_{nm}[\bar{\varepsilon}_{m,\mathbf{k}}(v_{\beta,\mathbf{k}})_{mn}+(v_{\beta,\mathbf{k}})_{mn}\bar{\varepsilon}_{n,\mathbf{k}}](\sigma_3)_{nn}(\sigma_3)_{mm}\frac{g(\bar{\varepsilon}_{n\mathbf{k}})-g(\bar{\varepsilon}_{m\mathbf{k}})}{(\bar{\varepsilon}_{n\mathbf{k}}-\bar{\varepsilon}_{m\mathbf{k}})^2}\nonumber\\
&&=\frac{1}{V}\sum_{m\neq n}\sum_\mathbf{k}-\frac{i}{2}(\sigma_3)_{nn}(\sigma_3)_{mm}\frac{[\bar{\varepsilon}_{n,\mathbf{k}}(v_{\beta,\mathbf{k}})_{nm}+(v_{\beta,\mathbf{k}})_{nm}\bar{\varepsilon}_{m,\mathbf{k}}](\theta_{\alpha,\mathbf{k}})_{mn}}{(\bar{\varepsilon}_{n\mathbf{k}}-\bar{\varepsilon}_{m\mathbf{k}})^2}g(\bar{\varepsilon}_{n\mathbf{k}})+c.c.\nonumber\\
&&=\frac{1}{V}\sum_{n\mathbf{k}}-(\Omega^\theta_{n,\mathbf{k}})^{\alpha}_\beta\bar{\varepsilon}_{n\mathbf{k}} g(\bar{\varepsilon}_{n\mathbf{k}})
-(m^\theta_{n,\mathbf{k}})^{\alpha}_\beta g(\bar{\varepsilon}_{n,\mathbf{k}}).
\end{eqnarray}

\subsection{Total response}
Adding the Kubo formula and dipole moment contributions together, the total response reads
\begin{eqnarray}
L^{\theta_\alpha}_{\beta}&&=S^{\theta_\alpha}_{\beta}+M^{\theta_\alpha}_{\beta}=\frac{1}{V}\sum_{n\mathbf{k}}(\Omega^\theta_{n,\mathbf{k}})^{\alpha}_\beta[-\bar{\varepsilon}_{n\mathbf{k}} g(\bar{\varepsilon}_{n\mathbf{k}})+ \int_0^{\bar{\varepsilon}_{n\mathbf{k}}}d\eta g(\eta)]=-\frac{1}{V} \sum_{n\mathbf{k}}(\Omega^\theta_{n,\mathbf{k}})^{\alpha}_\beta \int_0^{\bar{\varepsilon}_{n\mathbf{k}}}d\eta\eta \frac{dg(\eta)}{d\eta}\nonumber\\
&&=-\frac{1}{V} \sum_{n\mathbf{k}}(\Omega^\theta_{n,\mathbf{k}})^{\alpha}_\beta \tilde{c}_1(\bar{\varepsilon}_{n\mathbf{k}}),
\end{eqnarray}
where $\tilde{c}_1(x)=\int_0^xd\eta\eta \frac{dg(\eta)}{d\eta}$ with $g(\eta)=\frac{1}{e^{\beta\eta}-1}$. Using the relation $-g(-\eta)=1+g(\eta)$, we have $\tilde{c}(x)=\tilde{c}(-x)$. Therefore, the response function can be reduced to
\begin{eqnarray}
L^{\theta_\alpha}_{\beta}
&&=-\frac{1}{V} \sum_{n=1}^N\sum_\mathbf{k}[(\Omega^\theta_{n,\mathbf{k}})^{\alpha}_\beta \tilde{c}_1(\varepsilon_{n\mathbf{k}})+(\Omega^\theta_{n+N,\mathbf{k}})^{\alpha}_\beta \tilde{c}_1(-\varepsilon_{n,-\mathbf{k}})]=-\frac{1}{V} \sum_{n=1}^N\sum_\mathbf{k}[(\Omega^\theta_{n,\mathbf{k}})^{\alpha}_\beta +(\Omega^\theta_{n+N,-\mathbf{k}})^{\alpha}_\beta]\tilde{c}_1(\varepsilon_{n,\mathbf{k}})\nonumber\\
&&=-\frac{1}{V} \sum_{n=1}^N\sum_\mathbf{k}[(\Omega^\theta_{n,\mathbf{k}})^{\alpha}_\beta +(\Omega^\theta_{n+N,-\mathbf{k}})^{\alpha}_\beta][\tilde{c}_1(\varepsilon_{n,\mathbf{k}})-\int_0^{\infty}d\eta\eta \frac{dg(\eta)}{d\eta}]\nonumber\\
&&=-\frac{k_B T}{V} \sum_{n=1}^N\sum_\mathbf{k}[(\Omega^\theta_{n,\mathbf{k}})^{\alpha}_\beta +(\Omega^\theta_{n+N,-\mathbf{k}})^{\alpha}_\beta]c_1[g(\varepsilon_{n,\mathbf{k}})]
=-\frac{2k_B T}{V} \sum_{n=1}^N\sum_\mathbf{k}(\Omega^\theta_{n,\mathbf{k}})^{\alpha}_\beta c_1[g(\varepsilon_{n,\mathbf{k}})].
\end{eqnarray}
Here we used the properties of Berry curvature shown in Eq.~\eqref{Berrycurvature3} and \eqref{Berrycurvature4}, and the relation $-\int_{\varepsilon_n}^{\infty}\eta\frac{d g(\eta)}{d\eta}=\frac{1}{\beta}c_1[g(\varepsilon_n)]$.

\subsection{Property of Berry curvature}
Here we provide two useful properties of Berry curvature defined in \eqref{Berrycurvature1}. 

(1) Summation rule:
\begin{eqnarray}\label{Berrycurvature3}
\sum_{n=1}^{2N}(\Omega^\theta_{n,\mathbf{k}})^{\alpha}_\beta=\sum_{m\neq n}(\sigma_3)_{nn}(\sigma_3)_{mm}\text{Im}[\frac{\langle u_{n,\mathbf{k}}|\theta_\alpha| u_{m,\mathbf{k}}\rangle\langle u_{m,\mathbf{k}}|v_\beta|u_{n,\mathbf{k}}\rangle}{(\bar{\varepsilon}_{n,\mathbf{k}}-\bar{\varepsilon}_{m,\mathbf{k}})^2}+\frac{\langle u_{m,\mathbf{k}}|\theta_\alpha| u_{n,\mathbf{k}}\rangle\langle u_{n,\mathbf{k}}|v_\beta|u_{m,\mathbf{k}}\rangle}{(\bar{\varepsilon}_{n,\mathbf{k}}-\bar{\varepsilon}_{m,\mathbf{k}})^2}]=0.
\end{eqnarray}
In the middle step, we utilized the property that the band indices $m, n$ can be interchanged.\\

(2) Mapping between particle and hole space.

We note that the velocity operator $v_{\mathbf{k}}$ satisfies
\begin{eqnarray}
\sigma_1v_{\mathbf{k}}\sigma_1=\sigma_1\frac{\partial H_\mathbf{k}}{\partial \mathbf{k}}\sigma_1=-v^\ast_{-\mathbf{k}}.
\end{eqnarray}
At the same time, the particle-hole symmetry of the Hamiltonian enforces the relation 
\begin{eqnarray}
\sigma_1\theta_{\alpha,\mathbf{k}}\sigma_1=\theta_{\alpha,-\mathbf{k}}^\ast,
\end{eqnarray}
which is clearly satisfied when we consider the current and source term response for a given operator $\hat{O}$. Using the particle-hole symmetry property of the eigenstates and eigenvalues, we are able to show 
\begin{eqnarray}\label{Berrycurvature4}
(\Omega^\theta_{n,\mathbf{k}})^{\alpha}_\beta&&=\sum_{m(\neq n)}(\sigma_3)_{nn}(\sigma_3)_{mm}\frac{2\text{Im}(\langle u_{n,\mathbf{k}}|\theta_{\alpha,\mathbf{k}}|u_{m,\mathbf{k}}\rangle \langle u_{m,\mathbf{k}}|v_{\beta,\mathbf{k}}|u_{n,\mathbf{k}}\rangle)}{(\bar{\varepsilon}_{n,\mathbf{k}}-\bar{\varepsilon}_{m,\mathbf{k}})^2}\nonumber\\
&&=\sum_{m+N(\neq n+N)}(\sigma_3)_{n+N,n+N}(\sigma_3)_{m+N,m+N}\frac{2\text{Im}(\langle u^\ast_{n+N,-\mathbf{k}}|\sigma_1\theta_{\alpha,\mathbf{k}}\sigma_1|u^\ast_{m+N,-\mathbf{k}}\rangle \langle u^\ast_{m+N,-\mathbf{k}}|\sigma_1v_{\beta,\mathbf{k}}\sigma_1|u^\ast_{n+N,-\mathbf{k}}\rangle)}{(\bar{\varepsilon}_{n+N,-\mathbf{k}}-\bar{\varepsilon}_{m+N,-\mathbf{k}})^2}\nonumber\\
&&=\sum_{m(\neq n+N)}(\sigma_3)_{n+N,n+N}(\sigma_3)_{mm}\frac{2\text{Im}[(\langle u_{n+N,-\mathbf{k}}|\theta_{\alpha,-\mathbf{k}}|u_{m,-\mathbf{k}}\rangle)^\ast (\langle u_{m,-\mathbf{k}}|-v_{\beta,-\mathbf{k}}|u_{n+N,-\mathbf{k}}\rangle)^\ast]}{(\bar{\varepsilon}_{n+N,-\mathbf{k}}-\bar{\varepsilon}_{m,-\mathbf{k}})^2}\nonumber\\
&&=(\Omega^\theta_{n+N,-\mathbf{k}})^{\alpha}_\beta.
\end{eqnarray} 

\subsection{Calculation of Berry curvature}

\begin{figure}
    \centering
    \includegraphics[width=0.9\linewidth]{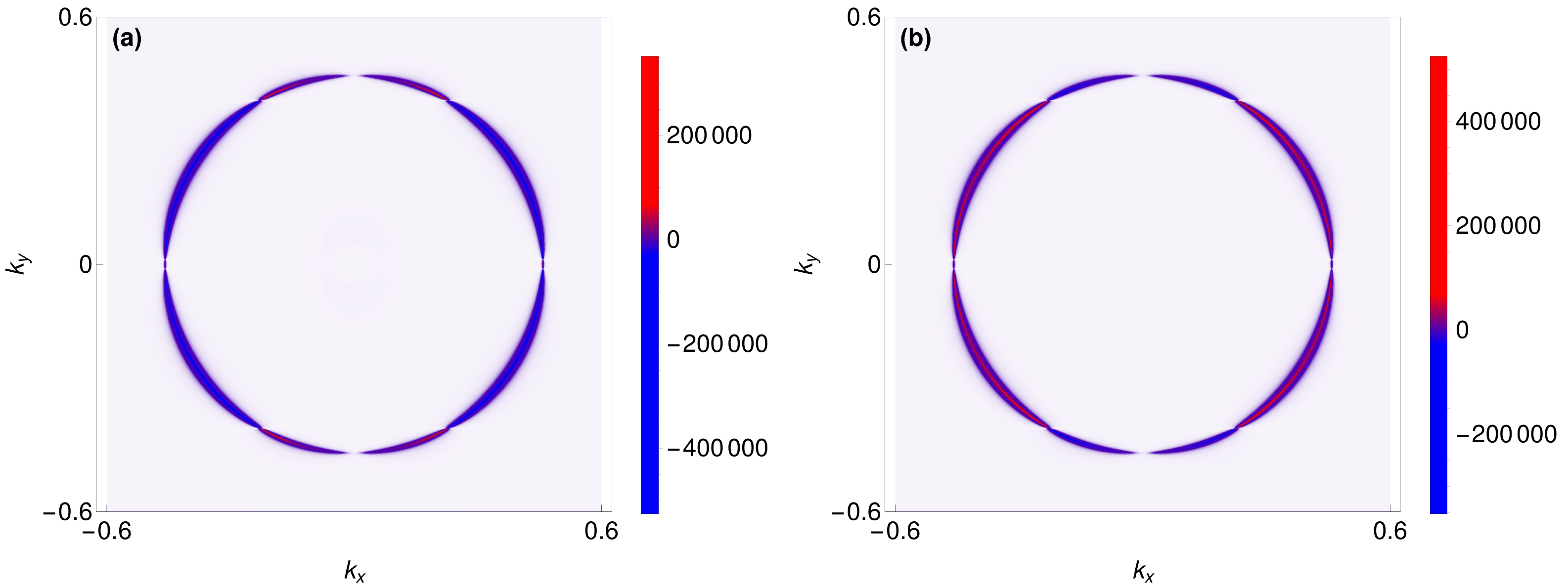}
    \caption{(Color online) Spin Berry curvature plots for kagome antiferromagnet KFe$_3$(OH)$_6$(SO$_4$)$_2$. (a): Middle energy band. (b): Lowest energy band.}
    \label{BC_Inset}
\end{figure}

In Fig.~\ref{BC_Inset}, we calculate the Berry curvature for kagome antiferromagnet KFe$_3$(OH)$_6$(SO$_4$)$_2$ using \begin{equation}
 (\Omega_{n,\mathbf{k}}^\theta)^{\alpha}_\beta=\sum_{m(\neq n)}(\sigma_3)_{nn}(\sigma_3)_{mm}\frac{2\text{Im}[(\theta_{\alpha,\mathbf{k}})_{nm}(v_{\beta,\mathbf{k}})_{mn}]}{(\bar{\varepsilon}_{n,\mathbf{k}}-\bar{\varepsilon}_{m,\mathbf{k}})^2}\nonumber
 \end{equation}

\section{Symmetry analysis on the spin Nernst tensor in kagome antiferromagnet}

We perform a detailed analysis on the effect of the (magnetic) point group on the Nernst response tensor. Suppose the Hamiltonian respects symmetry $g$ with matrix representation $U(g)$ for unitary operation and $U(g)\mathcal{K}$ for anti-unitary operation (containing time-reversal) with $\mathcal{K}$ being the complex conjugate operator. Here $U(g)$ corresponds to the point group operation on spin mode orbitals, which is a unitary matrix that satisfies $U(g)^\dagger=U(g)^T$. On the other hand, the point group symmetries don't mix particle and hole symmetry, such that $[\sigma_3, U(g)]=0$.
For the unitary case, we assume
\begin{eqnarray}
U(g)H_{\mathbf{k}}U^\dagger(g)=H_{M(g)\mathbf{k}},
\end{eqnarray}
where $M(g)$ is the matrix acting on momentum variables. We can deduce that
\begin{eqnarray}
|\psi_{n,M(g)\mathbf{k}}\rangle=U(g)|\psi_{n,\mathbf{k}}\rangle,\qquad \varepsilon_{M(g)\mathbf{k}}=\varepsilon_{\mathbf{k}}.
\end{eqnarray}
As a consequence, by inserting the symmetry operation in the matrix elements of an observable, we find 
\begin{eqnarray}
\langle\psi_{n,\mathbf{k}}|\hat{A}|\psi_{m,\mathbf{k}}\rangle=\langle\psi_{n,M(g)\mathbf{k}}|U(g)\hat{A}U(g)^\dagger|\psi_{m,M(g)\mathbf{k}}\rangle.
\end{eqnarray}
Similarly, for the anti-unitary case,
\begin{eqnarray}
U(g)H^\ast_{\mathbf{k}}U^\dagger(g)=H_{M(g)\mathbf{k}},
\end{eqnarray}
such that
\begin{eqnarray}
|\psi_{n,M(g)\mathbf{k}}\rangle=U(g)\mathcal{K}|\psi_{n,\mathbf{k}}\rangle,\qquad \varepsilon_{M(g)\mathbf{k}}=\varepsilon_{\mathbf{k}}.
\end{eqnarray}
These relations will lead to
\begin{eqnarray}
\langle\psi_{n,\mathbf{k}}|\hat{A}|\psi_{m,\mathbf{k}}\rangle=\langle\psi_{n,M(g)\mathbf{k}}|U(g)\hat{A}U(g)^\dagger|\psi_{m,M(g)\mathbf{k}}\rangle^\ast.
\end{eqnarray}
If the operator $\hat{A}$ satisfies
\begin{eqnarray}
U(g)\hat{A}_iU(g)^\dagger=\sum_jR(g)_{ij}\hat{A}_j,
\end{eqnarray}
and we combine this with the element's symmetry relation, we can obtain a transformation relation for the spin Nernst response coefficient 
\begin{eqnarray}\label{Symmetryconstrain}
\alpha^\gamma_{\lambda\beta}=\pm R^s(g)_{\gamma i}R^v(g)_{\lambda j}R^v(g)_{\beta k}\alpha^i_{jk},
\end{eqnarray} 
where the plus and minus sign correspond to unitary and anti-unitary symmetry, and $R^{s/v}(g)$ stands for the transformation matrix for the spin and velocity operator, respectively.  Moreover, suppose the involved non-magnetic point group symmetry $U(g)$ corresponds to a spatial operation with matrix form $R$ in Cartesian coordinates. If $\hat{g}$ is a unitary operation,
\begin{eqnarray}\label{Unitary}
R^s(g)=\text{det}(R)R,\qquad R^v(g)=R.
\end{eqnarray}
For anti-unitary operation,
\begin{eqnarray}\label{Antiunitary}
R^s(g)=-\text{det}(R)R,\qquad R^v(g)=-R.
\end{eqnarray}
Plugging Eq.~\eqref{Unitary}, \eqref{Antiunitary} into Eq.~\eqref{Symmetryconstrain}, we find
\begin{eqnarray}\label{Symmetryconstrain1}
\alpha^\gamma_{\lambda\beta}=\text{det}(R)R_{\gamma \gamma^\prime}R_{\lambda \lambda^\prime}R_{\beta \beta^\prime}\alpha^{\gamma^\prime}_{\lambda^\prime\beta^\prime}
\end{eqnarray}

In the kagome AF, we focus on two symmetries of the system: the mirror reflection with respect to the $y-z$ plane plus time-reversal $\hat{g}_1=\mathcal{M}_{yz}\mathcal{T}$, and the threefold rotation about the $z$ axis $\hat{g}_2=\mathcal{C}_{3z}$. It is straightforward to obtain the matrix representation in Cartesian coordinates of these two symmetry operations
\begin{eqnarray}
R(g_1)=\left(\begin{array}{ccc}
-1&0&0\\
0&1&0\\
0&0&1
\end{array}\right),\qquad
R(g_2)=\left(\begin{array}{ccc}
-\frac{1}{2}&-\frac{\sqrt{3}}{2}&0\\
\frac{\sqrt{3}}{2}&-\frac{1}{2}&0\\
0&0&1
\end{array}\right).
\end{eqnarray}
By applying these symmetries to Eq.~\eqref{Symmetryconstrain1}, the spin Nernst response tensor (only considering in-plane driven response) can be fixed to
\begin{eqnarray}
\alpha^{x}=\left(\begin{array}{cc}
-\alpha^{y}_{yx}&0\\
0&\alpha^{y}_{yx}
\end{array}\right),\qquad
\alpha^{y}=\left(\begin{array}{cc}
0&\alpha^{y}_{yx}\\
\alpha^{y}_{yx}&0
\end{array}\right),\qquad
\alpha^{z}=\left(\begin{array}{cc}
0&-\alpha^{z}_{yx}\\
\alpha^{z}_{yx}&0
\end{array}\right),
\end{eqnarray}
where $\alpha^y_{yx}$ and $\alpha^z_{yx}$ correspond $\alpha_1$, $\alpha_2$ in the main text individually. Here we comment that the $g_1$ symmetry can also be replaced by $\mathcal{C}_{2x}\mathcal{T}$, the twofold rotation about the $x$ axis plus time reversal.

\end{widetext}

\end{document}